\documentclass[sigconf,nonacm,review=false,timestamp=false]{acmart}

\usepackage{multirow}
\usepackage{siunitx}
\usepackage{comment}
\usepackage{subcaption} % For subfigures
\usepackage{algorithm}
\usepackage{listings}
\usepackage{makecell}
\usepackage{geometry}
\usepackage{libertine}
\usepackage{xcolor}

\definecolor{ColorJSON}{HTML}{2596be}

\AtBeginDocument{%
  \providecommand\BibTeX{{%
    \normalfont B\kern-0.5em{\scshape i\kern-0.25em b}\kern-0.8em\TeX}}}

\graphicspath{{./images/}} 

\lstdefinelanguage{hjson}{
    keywords=[1]{module, endmodule, input, output, wire, reg, always, always_comb, always_ff, if, else, ‘ifdef, ‘endif, case, endcase, begin, end, assign, initial},
    keywords=[2]{logic, int, bit, logic, void, return, function, endfunction, task, endtask, parameter, localparam, typedef},
    keywordstyle=[1]\color{blue},   % Color of control flow keywords
    keywordstyle=[2]\color{purple}, % Color of data type keywords
    sensitive=true,
    comment=[l]{//},                % Line comment style
    morecomment=[s]{/*}{*/},        % Block comment style
    commentstyle=\color{gray}\ttfamily, % Comment color and style
    morestring=[b]",                % Strings style
    stringstyle=\color{ColorJSON}\ttfamily, % String color and style
    basicstyle=\ttfamily\footnotesize,  % Basic font style for code
    numberstyle=\tiny\color{gray},   % Line number style
    stepnumber=1,                   % Numbering interval
    numbersep=5pt,                  % Space between line numbers and code
    tabsize=4,                      % Tabs size
    showspaces=false,               % Do not display spaces
    showstringspaces=false,         % Do not display string spaces
    breaklines=true,                % Enable line break
    frame=single,                   % Add a frame around the code
    backgroundcolor=\color{lightgray!20},  % Background color
}

\begin{document}

%%
%% The "title" command has an optional parameter,
%% allowing the author to define a "short title" to be used in page headers.
\title[Just TestIt!]{Just TestIt!\\An SBST Approach To Automate System-Integration Testing}

%%
%% The "author" command and its associated commands are used to define
%% the authors and their affiliations.
%% Of note is the shared affiliation of the first two authors, and the
%% "authornote" and "authornotemark" commands
%% used to denote shared contribution to the research.
%\author{Anonymous authors}
\author{Tommaso Terzano}
\affiliation{%
  \institution{Politecnico di Torino}
  \city{Torino}
  \country{Italy}}
\email{tommaso.terzano@polito.it}

\author{Luigi Giuffrida}
\affiliation{%
  \institution{Politecnico di Torino}
  \city{Torino}
  \country{Italy}}
\email{luigi.giuffrida@polito.it}

\author{Juan Sapriza}
\affiliation{%
  \institution{École Polytechnique Fédérale de
 Lausanne}
  \city{Lausanne}
  \country{Switzerland}}
\email{juan.sapriza@epfl.ch}

\author{Pasquale Davide Schiavone}
\affiliation{%
  \institution{École Polytechnique Fédérale de
 Lausanne}
  \city{Lausanne}
  \country{Switzerland}}
\email{davide.schiavone@epfl.ch}

\author{Guido Masera}
\affiliation{%
  \institution{Politecnico di Torino}
  \city{Torino}
  \country{Italy}}
\email{guido.masera@polito.it}

\author{David Atienza}
\affiliation{%
  \institution{École Polytechnique Fédérale de
 Lausanne}
  \city{Lausanne}
  \country{Switzerland}}
\email{david.atienza@epfl.ch}

\author{Luciano Lavagno}
\affiliation{%
  \institution{Politecnico di Torino}
  \city{Torino}
  \country{Italy}}
\email{luciano.lavagno@polito.it}

\author{Maurizio Martina}
\affiliation{%
  \institution{Politecnico di Torino}
  \city{Torino}
  \country{Italy}}
\email{maurizio.martina@polito.it}

%%
%% By default, the full list of authors will be used in the page
%% headers. Often, this list is too long, and will overlap
%% other information printed in the page headers. This command allows
%% the author to define a more concise list
%% of authors' names for this purpose.
%\renewcommand{\shortauthors}{Anonymous, et al.}
\renewcommand{\shortauthors}{Terzano, et al.}

%%
%% The abstract is a short summary of the work to be presented in the
%% article.
\begin{abstract}
\textbf{This paper introduces TestIt, an open-source Python package designed to automate full-system integration testing using a Software-Based Self-Test (SBST) approach. By dynamically generating test vectors and golden references, TestIt significantly reduces development time and complexity while supporting both simulation and FPGA environments. Its flexible design positions TestIt as a key enabler for the widespread adoption of CI/CD methodologies in open-source RTL development.}
\textbf{A case study on the X-HEEP RISC-V microcontroller (MCU), which integrates a custom accelerator, showcases TestIt’s ability to detect hardware and software faults that traditional formal methods may overlook. Furthermore, the case study highlights how TestIt can be leveraged to characterize system performance with minimal effort. By automating testing on the PYNQ-Z2 FPGA development board, we achieved a $11\times$ speed-up with respect to RTL simulations.}
\end{abstract}

%%
%% The code below is generated by the tool at http://dl.acm.org/ccs.cfm.
%% Please copy and paste the code instead of the example below.
%%
% \begin{CCSXML}
% <ccs2012>
%    <concept>
%        <concept_id>10010583.10010737.10010739</concept_id>
%        <concept_desc>Hardware~Board- and system-level test</concept_desc>
%        <concept_significance>500</concept_significance>
%        </concept>
%    <concept>
%        <concept_id>10010583.10010717.10010721</concept_id>
%        <concept_desc>Hardware~Functional verification</concept_desc>
%        <concept_significance>500</concept_significance>
%        </concept>
%  </ccs2012>
% \end{CCSXML}


%%
%% This command processes the author and affiliation and title
%% information and builds the first part of the formatted document.
\maketitle
\section{Introduction}
Verifying the functionality of a digital integrated circuit (IC) is a complex yet critical step in its development. While formal verification methods are highly effective for targeting individual components, their computational complexity increases exponentially, making them impractical for testing large-scale systems, like heterogeneous MCUs. Additionally, these methods are limited to hardware verification, leaving software layers untested.

Integration testing can be introduced as an additional step in the pre-production phase to alleviate formal verification limitations. This approach incrementally validates the entire system by assessing interactions between previously verified components.
Moreover, integration testing can be executed on hardware platforms such as FPGAs at operational speed. This allows for the execution of long tests and real-scenario end-to-end applications, which can include interactions with external components (such as external memories, ADCs, DACs, etc.) as well as the whole data-flow digital-processing chain. 
Once developed, integration tests can also be used for post-production validation to ensure that the final product behaves correctly.

SBST techniques further enhance flexibility while eliminating the need for costly Automated Test Equipment (ATE) \cite{ruosposbst}.

While several solutions have been proposed to optimize formal verification phases, as described in the next section, no open-source tool currently targets integration testing.

This paper presents \textit{TestIt}\footnote{GitHub repository URL: https://github.com/vlsi-lab/TestIt}, a Python package that provides a highly flexible SBST-based solution for developing full-system integration tests. \textit{TestIt} supports both simulation and FPGA targets, automating the entire workflow: from generating random datasets to interfacing with the FPGA development board.

\section{Related works}

Automating testing and verification is an increasingly important aspect of the design of modern complex Systems-on-Chip (SoCs). Indeed, numerous frameworks have been developed to automate the generation of test-benches and formal verification environments. 

However, they primarily focus on unit testing or platform development, leaving system-level integration testing largely unaddressed.  
Furthermore, existing solutions are unable to comprehensively test interactions between the hardware platform and the software stack, which is essential for validating real-world functionality.

\begin{table}[t]
\centering
\caption{Comparison of the analyzed frameworks}
\label{tab:comparison}
\begin{tabular}{lcccc}
\toprule
\textbf{Feature} & \textbf{AutoSVA} & \textbf{Cocotb} & \textbf{Renode} & \textbf{TestIt} \\
\midrule
High-Level Models       & \textcolor{red}{$\times$} & \textcolor{red}{$\times$} & \textcolor{green}{$\checkmark$} & \textcolor{green}{$\checkmark$} \\ \hline
RTL-Level Models        & \textcolor{green}{$\checkmark$} & \textcolor{green}{$\checkmark$} & \textcolor{red}{$\times$} & \textcolor{green}{$\checkmark$}\\ \hline
\makecell{System-Level \\ Integration} & \textcolor{red}{$\times$} & \textcolor{red}{$\times$} & \textcolor{green}{$\checkmark$} & \textcolor{green}{$\checkmark$}\\ \hline
\makecell{Automated \\ Randomized Tests}           & \textcolor{green}{$\checkmark$} & \textcolor{green}{$\checkmark$} & \textcolor{red}{$\times$} & \textcolor{green}{$\checkmark$}\\ \hline
Simulated Testing       & \textcolor{green}{$\checkmark$} & \textcolor{green}{$\checkmark$} & \textcolor{green}{$\checkmark$} & \textcolor{green}{$\checkmark$}\\ \hline
FPGA Testing            & \textcolor{red}{$\times$}  & \textcolor{red}{$\times$}  & \textcolor{red}{$\times$} & \textcolor{green}{$\checkmark$}\\ \hline
SW Stack Testing        & \textcolor{red}{$\times$} & \textcolor{red}{$\times$} & \textcolor{red}{$\times$} & \textcolor{green}{$\checkmark$}\\
\bottomrule
\end{tabular}
\end{table}
\vspace{-10pt}
\textit{Renode} \cite{renode} allows users to assemble virtual SoCs using modular building blocks, including ARM and RISC-V CPUs, as well as various communication buses and interfaces. However, \textit{Renode} operates at a quite high level of abstraction; it lacks support for automated test execution, and it works exclusively in a simulated environment.

\textit{AutoSVA} \cite{orenes2021autosva} automates formal testbench generation for unit-level RTL verification. 
While reducing development effort compared to UVM, it still requires additional work in RTL design. Its primary focus is formal verification, which, despite optimizations, remains highly time-consuming for large-scale systems integration. 
Additionally, \textit{AutoSVA} relies on an extra software layer to perform the verification process, currently limited to \textit{SymbiYosys} \cite{SymbiYosys}.

Finally, \textit{cocotb} \cite{cocotb} is a testbench environment for verifying RTL designs similar to the UVM approach, but using Python. While \textit{cocotb} can reduce the overhead of test creation, it does not automate the test flow, it doesn't enable software-layer testing, and it's limited to simulated environments.

\autoref{tab:comparison} presents a visual comparison of the analyzed frameworks, highlighting how \textit{TestIt} addresses the gaps identified in state-of-the-art alternatives.

\section{Overview}

\begin{figure}[t]
    \centering
    \includegraphics[width=\linewidth]{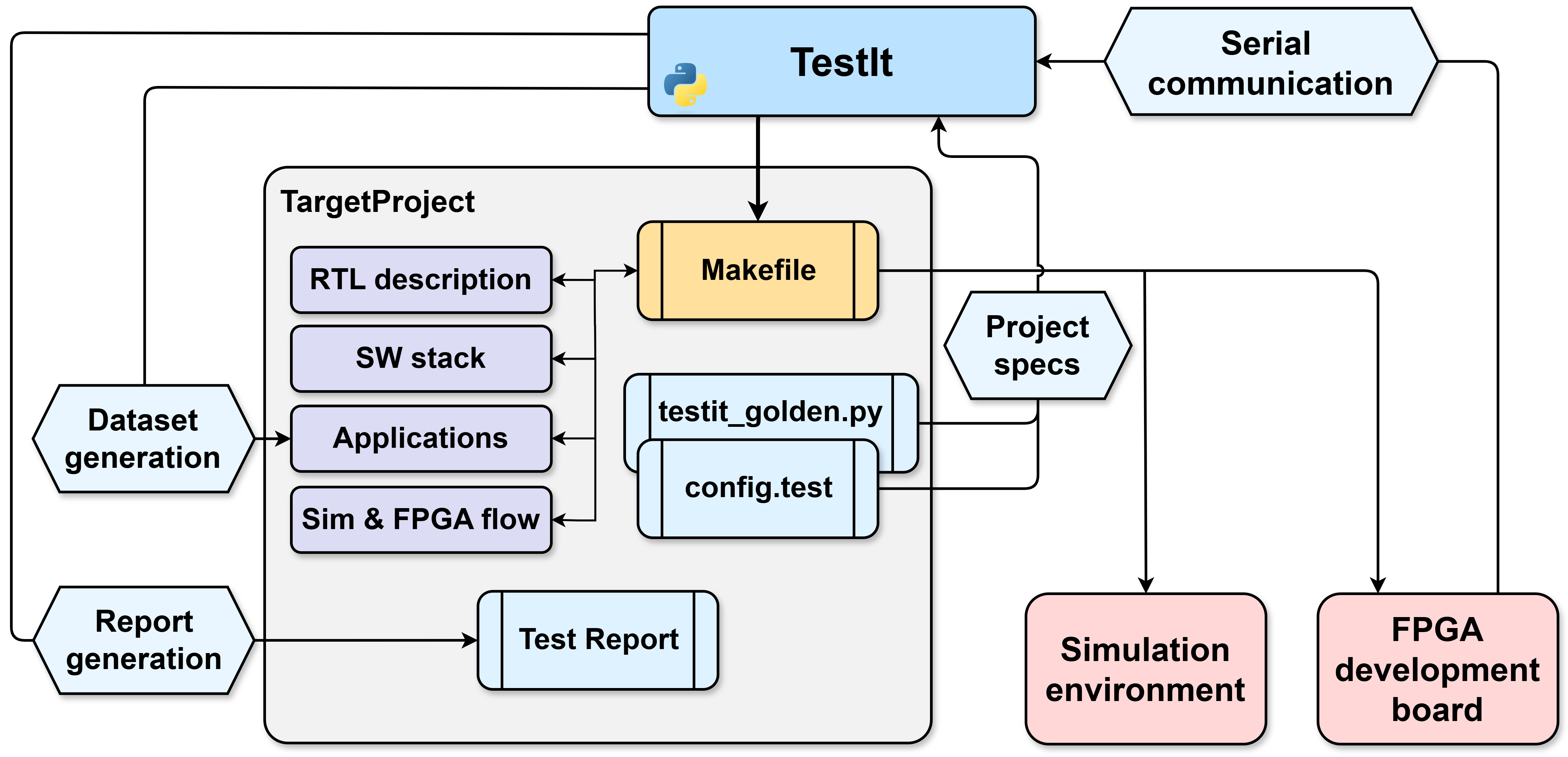}
    \caption{Structure of a TestIt environment}
    \label{fig:testit_scheme}
% \vspace{-15pt}
\end{figure}

\textit{TestIt} is a Python package that implements a command-line application for executing comprehensive integration test campaigns using an SBST approach. It leverages software-driven testing to verify both the integration of hardware components and the correct operation of the software stack, from HAL (Hardware Abstraction Layer) drivers to applications. 

To ensure sufficient randomness, \textit{TestIt} dynamically generates a unique input dataset for each test iteration, along with a corresponding golden reference dataset. These datasets are written by the tool into a C-code source and header files pair.  These files are linked during compilation, allowing the application to validate the correctness of the System-Under-Test (SUT), without relying on external dependencies.

\subsection{Requirements}

To function properly, \textit{TestIt} requires a few key components:  
\begin{itemize}  
    \item A \textbf{Target RTL Project} with a complete and functional development flow, including software compilation, model synthesis, simulation, and debugger support.

    \item A set of predefined \textbf{Make targets} in the project's Makefile, enabling \textit{TestIt} to interact with the existing workflow.

    \item A \textbf{configuration file} that defines the project's structure and test parameters.  

    \item A \textbf{Python module}, developed by the test engineers, containing the functions required by \textit{TestIt} to generate the golden datasets.
\end{itemize}  

\paragraph{\textbf{Makefile Targets}}

\textit{TestIt} relies on eight predefined Makefile targets that must be present in the RTL project. They provide maximum flexibility, allowing developers to implement each one according to their specific project characteristics.

The required Makefile targets are:  
\begin{description}  
    \item[\texttt{sw-sim [app]}]: Compiles software applications for simulation environments. Accepts an \texttt{app} argument.  
    \item[\texttt{sw-fpga [app]}]: Compiles software applications for FPGA development boards. Accepts the same "\texttt{app}" argument.  
    \item[\texttt{sim-build [tool]}]: Builds the simulation model, with a "\texttt{tool}" argument.  
    \item[\texttt{sim-run [tool]}]: Sets up and runs the simulation, also with a "\texttt{tool}" argument.  
    \item[\texttt{fpga-build [target]}]: Builds an FPGA model, which can be implemented as a bitstream. Accepts a "\texttt{target}" argument.  
    \item[\texttt{fpga-load [target]}]: Loads the FPGA model onto the FPGA development board. Accepts the same "\texttt{target}" argument. 
    \item[\texttt{gdb-setup}]: Sets up the GDB debugger of choice.  
    \item[\texttt{deb-setup}]: Configures the preferred debugging tool, such as OpenOCD.  
\end{description}  

\paragraph{\textbf{Test Configuration Files}}  

To run \textit{TestIt}, two configuration files are required, which can be generated in the working directory using the command "\texttt{testit setup}", as described in the next paragraph. These files are:
\begin{description}
    \item[\texttt{config.test}]: An HJSON file containing all necessary information about the target project, including the test descriptions.  
    \item[\texttt{testit\_golden.py}]: A Python module that contains the functions used to generate the golden result dataset, which the application utilizes to compare its outputs and verify correct behavior.  
\end{description}  

The \texttt{config.test} file consists of three main fields: \textbf{target}, \textbf{report}, and \textbf{test}. Each field provides essential configuration details for \textit{TestIt}.
\subsubsection*{\textbf{Target Field [\ref{cfg_target}]}}
This specifies details about the test environment, including the test platform’s name and type, serial connection settings (for FPGA boards), the number of random iterations, and the output file directory when using a simulation tool.
\subsubsection*{\textbf{Report Field [\ref{cfg_target}]}}
This field specifies the directory where \textit{TestIt} stores the test report and HJSON test data.

% \vspace{-12pt}
\begin{figure}[t]
\begin{lstlisting}[language=hjson, label=cfg_target, caption={Example \texttt{config.test} - Target and Report Field}]
  target: {
    name: "pynq-z2"
    type: "fpga"
    usbPort: 2 
    baudrate: 9600
    iterations: 10
    outputFile: "path/to/sim/dump"
  }
  report: {
    dir: "path/to/report/folder"
  }
\end{lstlisting}  
\end{figure}
% \vspace{-15pt}

\subsubsection*{\textbf{Test Field [\ref{cfg_test}]}}
This field defines all tests executed within a single iteration. Each entry includes:  
\begin{itemize}
    \item The application name and its directory.
    \item The \texttt{.c} and \texttt{.h} files storing input and golden datasets.
    \item The regular expression used by the SUT to communicate test results to the host, either via serial communication for FPGA boards or file dumping in a simulation environment.
    \item A list of test parameters with fixed values or ranges.
    \item Input and output datasets, specifying data type, value range, and dimensions, which can be parameter dependent.
    \item The golden function used to generate reference values.
\end{itemize}

% \vspace{-15pt}
\begin{figure}[t]
\begin{lstlisting}[language=hjson, label=cfg_test, caption={Example \texttt{config.test} - Test Field}]
  test: [
    {
      appName: "application_name"
      dir: "path/to/app"
      genFilesName: "test_data"
      outputFormat: "(\\d+):(\\d+):(\\d+)" 
      outputTags: ["TestID", "Cycles", "Outcome"]
      parameters: [
        {
          name: "SIZE"
          value: [4, 10]
          step: 2
        }
      ]
      inputDataset: [
        { 
          name: "input_matrix"
          dataType: "uint8_t"
          valueRange: [0, 255]
          dimensions: ["SIZE", "SIZE"]
        }
      ]
      outputDataset: [
        { 
          name: "output_matrix"
          dataType: "uint8_t"
        }
      ]
      goldenResultFunction: {
        name: "softmax"
      }
    }
  ]
\end{lstlisting}  
\end{figure}
% \vspace{-15pt}

Each test can define parameters with a name, value, and optionally a range with a step size. If a parameter is defined as a range, \textit{TestIt} selects a value within it for each iteration. These parameters are included in the \texttt{.h} file and passed as arguments to the golden function in \texttt{testit\_golden.py}, which can use them for the computation of reference values, if needed.  

For each dataset, it is possible to specify the name of the C array storing the data, its data type, the range of values for random generation, and its dimensions. If the dimensions depend on a parameter, \textit{TestIt} parses the corresponding parameter value for every matching dimension.
\subsection{TestIt Commands}

Following the design criterion of simplicity, TestIt needs just three commands to run the test campaign.

\paragraph{\textbf{testit setup}}

This command checks for the presence of the required files, \texttt{config.test} and \texttt{testit\_golden.py}, in the working directory. 
If these files are not found, it automatically generates fully commented templates that can be easily modified and tailored to the specific needs of the target project.

\paragraph{\textbf{testit run}}

This is the core command of \textit{TestIt}, enabling the execution of an integration test campaign as defined in \texttt{config.test}.  

The process begins with initial checks on the target project's Makefile and \texttt{config.test}. 
Afterward, the simulation or FPGA model is built. If needed, this step can be skipped by using the "\texttt{\texttt{-}{-}nobuild}" flag.  

In the case of FPGA-based testing, the model is loaded onto the development board, the serial connection is initialized, and the debugger is set up.

The actual testing process then begins. Each iteration starts with dataset generation, which can be performed in two ways:  
\begin{itemize}
    \item By default, \textit{TestIt} uses the "\texttt{iterations}" parameter in the \texttt{config.test} file. For each iteration, it selects random parameter values if required.  
    \item Alternatively, if the command "\texttt{testit run}" is executed with the "\texttt{\texttt{-}{-}sweep}" flag, \textit{TestIt} systematically tests all possible parameter value combinations, incrementing values according to the "\texttt{step}" parameter.  
\end{itemize}  

Following the SBST approach, the application itself processes the input dataset and compares it with the golden result to verify the functionality of the SUT.

\textit{TestIt} parses the output of the test using the regular expressions specified in "\texttt{outputFormat}" and "\texttt{outputTags}" parameters from \texttt{config.test}. This extracted information is then processed and stored in a JSON database within the report directory.

This approach was chosen over alternatives such as memory reads via \textit{gdb} due to its greater flexibility. It allows test engineers to specify, for each individual test application, which data should be acquired. In some cases, this may be limited to the test ID and result, but the regular expression can be extended to include additional details such as execution cycles, the number of detected errors, and other relevant metrics.  

\paragraph{\textbf{testit report}}

Finally, \textit{TestIt} can generate a summary report containing the data acquired during the test campaign. The report can be displayed directly in the terminal sorted by any metric extracted from the regular expression using the "\texttt{\texttt{-}{-}sort\_key [KEY]}" flag, with the option to sort in descending order by specifying "\texttt{\texttt{-}{-}descending}".  

\section{Use Case: Custom Accelerator Integration in X-HEEP MCU}

As previously discussed, \textit{TestIt} has been developed primarily for system-level integration testing.  
To validate this approach and demonstrate its potential, we present an integration test campaign conducted on X-HEEP \cite{xheep}, an open-source 32-bit RISC-V MCU. 

In this scenario, X-HEEP has been extended with a Smart Peripheral Controller (SPC) accelerator that optimizes the \textit{im2col} reshaping transformation, enabling convolutions to be performed as single matrix multiplications \cite{thesis}.  
This accelerator is tightly integrated with the X-HEEP's DMA to perform 2D transactions, which are called sequentially by the im2col accelerator to perform the final transformation. Due to its tight integration within the X-HEEP platform, it serves as a prime example of the need for system-level integration testing.

% \vspace{-10pt}
\subsection{Performance Characterization}
Performance characterization is crucial for IC design because it enables designers to find optimal trade-offs. While simulations can be time-consuming, \textit{TestIt}'s FPGA support and automation features allow efficient characterization of RTL designs.

\begin{figure}[t]
% \vspace{-10pt}
\centering
\includegraphics[width=\linewidth]{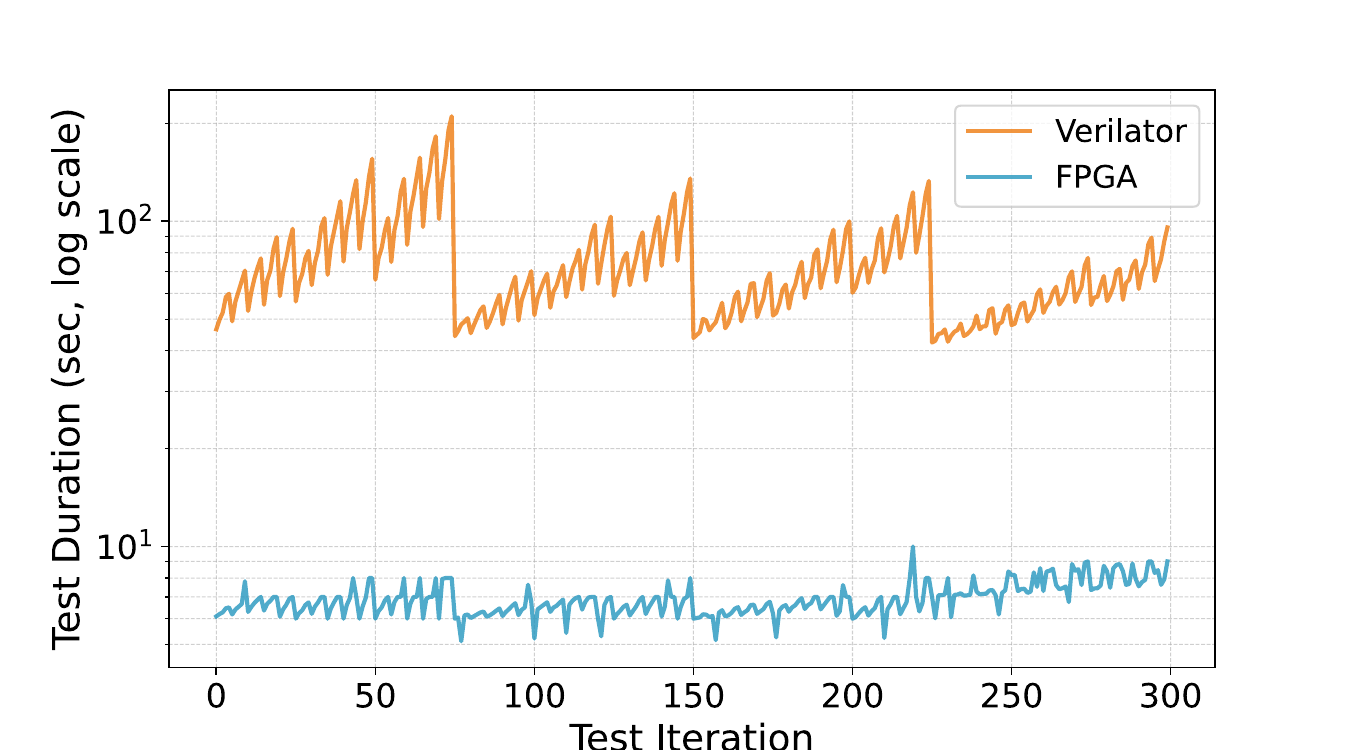}
\caption{Test duration in seconds}
\label{fig:test_duration}
% \vspace{-5pt}
\end{figure}

To demonstrate the speedup that FPGA-based testing can achieve, we carried out a 300 iteration test campaign both on the PYNQ-Z2 FPGA development board and performing Verilator simulations, targeting the im2col SPC+X-HEEP system. 
The test application was modified to take advantage of the on-board timer to record timestamps before and after each test. Their difference was then transmitted to the host device using the previously mentioned regular expression mechanism, thus enabling performance characterization. The FPGA-based campaign took around 0:31 hours to complete, while the Verilator-based one took 6:07 hours, a \textbf{$11\times$} increase in test time.
\autoref{fig:test_duration} graphically compares the two approaches by showing the duration of each test iteration in seconds on a logarithmic scale.

% \vspace{-10pt}
\subsection{Hardware fault}

To evaluate the effectiveness of our approach, we injected a hardware fault into the SPC, one that could only be detected through system-level testing.
Specifically, we modified the controller responsible for managing the communication between the accelerator and X-HEEP’s peripheral interface.
The used protocol implements a simple request-response handshake, where the controller must wait for the response valid signal before de-asserting the request valid signal. Our fault removed this synchronization mechanism, potentially disrupting the entire functionality of the SPC.

In a standard \textit{UVM} environment, idealized external interfaces cause the response valid signal to consistently appear one cycle later, rendering synchronization issues undetectable even in full system-level simulations, whereas FPGA-based testing introduces real-world effects that can reveal such flaws.

% \vspace{-10pt}
\subsection{Software Fault}

\begin{figure}[t]
\centering
\includegraphics[width=\linewidth]{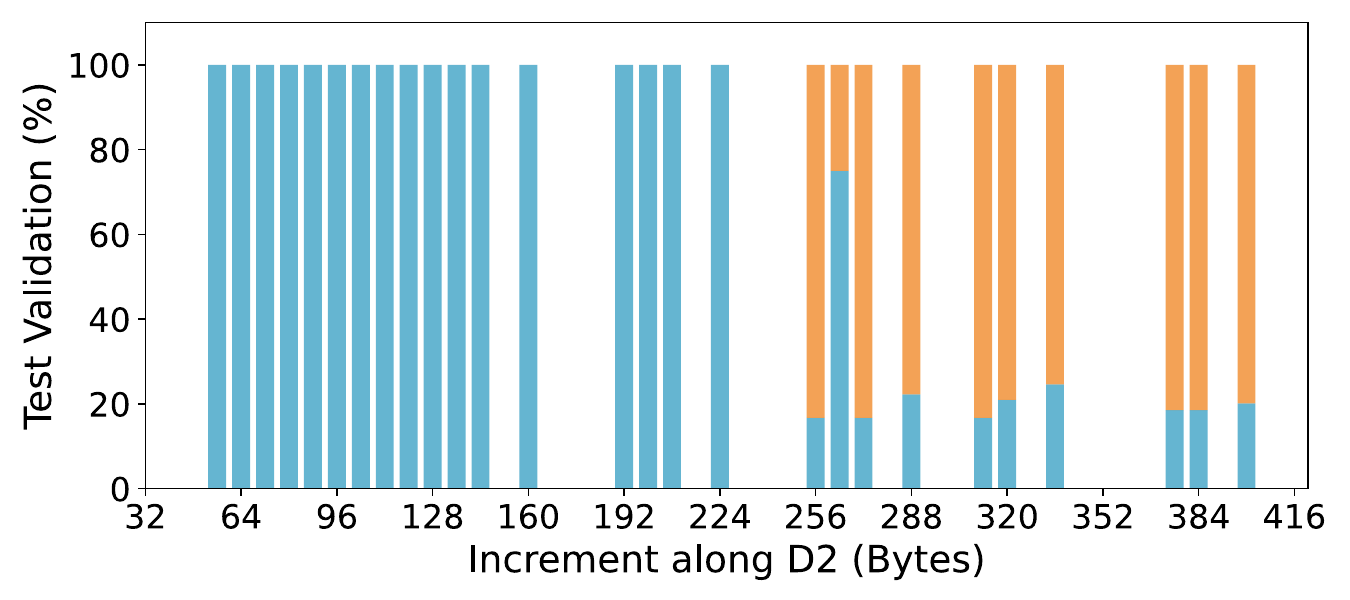}
\caption{Integration test results with the DMA HAL fault}
\label{fig:im2col_plot_all}
\end{figure}

To assess \textit{TestIt}'s ability to validate the software stack, we conducted a test campaign on the im2col transformation using three implementations: a C-based algorithm, a DMA-optimized version, and the im2col SPC.
To evaluate the fault detection capabilities of the tool, we introduced an error in the DMA HAL function \texttt{get\_increment\_b\_1D()}. Specifically, a \texttt{uint8\_t} variable was incorrectly used instead of a \texttt{uint32\_t} for computing byte increments, leading to an overflow error in sufficiently large transactions.
The fault appears only in the DMA-based test, while the \textit{im2col} SPC functions correctly. This distinction enables test engineers to quickly identify the issue as software-related rather than hardware-related.
\autoref{fig:im2col_plot_all} presents the results of this testing campaign, illustrating how the size of the modified variable impacts test outcomes. Notably, issues arise when increments exceed 256 words, triggering the overflow error.

% \vspace{-5pt}
% \vspace{-15pt}

\section{Future works}

Continuous Integration and Continuous Deployment (CI/CD) is a widely adopted methodology that aims at simplifying and accelerating development cycles. 
It consists of automatically integrating code changes into a shared repository, running extensive automated tests, and deploying updates with minimal manual intervention. 
While CI/CD is a popular practice in software engineering, its adoption in hardware design remains limited.

As a future development, we propose leveraging \textit{TestIt}'s capabilities to implement an FPGA-based CI/CD system for RTL projects. 
This would involve integrating a \textit{TestIt}-based environment into the development workflow using a self-hosted GitHub Action. Upon each git push to the repository, \textit{TestIt} would automate FPGA synthesis, model loading, and test execution, streamlining the integration and validation of new features.
Therefore, \textit{TestIt} could serve as a critical enabler in integrating CI/CD methodologies within open-source RTL development, especially when combined with unit-level formal verification.
In conclusion, this work represents a significant step forward in the broader adoption of agile practices \cite{agileHWPATTERSON} in hardware design.

%%
%% The acknowledgments section is defined using the "acks" environment
%% (and NOT an unnumbered section). This ensures the proper
%% identification of the section in the article metadata, and the
%% consistent spelling of the heading.
%\begin{acks}
%Acknowledgements go here. Delete enclosing begin/end markers if there are no acknowledgements.
%\end{acks}

%%
%% The next two lines define the bibliography style to be used, and
%% the bibliography file.
% \vspace{-5pt}
\begin{acks}
This work is part of the project NODES which has received funding from the MI\_JQ — M4C21 5 of PNRR funded by the European Union - NextGenerationEU (Grant agreement no. ECSOOOOOOB6). Also, this work was supported in part by the the Swiss NSF Edge-Companions project (GA No. 10002812); in part by the EC H2020 FVLLMONTI Project under Grant 101016776; in part by the ACCESS—AI Chip Center for Emerging Smart Systems, sponsored by InnoHK funding, Hong Kong, SAR; and in part by the Swiss State Secretariat for Education, Research, and Innovation (SERI) through the SwissChips Research Project.
\end{acks}

% \vspace{-5pt}
\bibliographystyle{ACM-Reference-Format}
\bibliography{references.bib}

\end{document}